\documentclass[conference]{IEEEtran}
\usepackage{cite}

\ifCLASSINFOpdf
\else
\fi
\usepackage[cmex10]{amsmath}
\usepackage{algorithm}
\usepackage{algorithmic}
\usepackage{multirow}

\usepackage{array}

\usepackage{bbding}
\usepackage{multirow}
\usepackage{booktabs}
\usepackage{url}

\usepackage{graphicx}

\usepackage[tight,footnotesize]{subfigure}

\usepackage{stfloats}

\usepackage{url,fancyhdr}

\begin{document}
%
\title{Large-scale Artificial Neural Network:\\ MapReduce-based Deep Learning}

\author{Kairan Sun, Xu Wei, Gengtao Jia, Risheng Wang, and Ruizhi Li \\
\IEEEauthorblockA{Department of Electrical and Computer Engineering \\
University of Florida\\
Gainesville, Florida 32611}}

\pagestyle{fancy}
\fancyhead[LO]{\small Project Report for Cloud Computing and Storage. Dept. of Electrical and Computer Engineering, University of Florida, \today}


%


\maketitle

\begin{abstract}
Faced with continuously increasing scale of data, original back-propagation neural network based machine learning algorithm presents two non-trivial challenges: huge amount of data makes it difficult to maintain both efficiency and accuracy; redundant data aggravates the system workload. This project is mainly focused on the solution to the issues above, combining deep learning algorithm with cloud computing platform to deal with large-scale data. A MapReduce-based handwriting character recognizer will be designed in this project to verify the efficiency improvement this mechanism will achieve on training and practical large-scale data. Careful discussion and experiment will be developed to illustrate how deep learning algorithm works to train handwritten digits data, how MapReduce is implemented on deep learning neural network, and why this combination accelerates computation. Besides performance, the scalability and robustness will be mentioned in this report as well. Our system comes with two demonstration software that visually illustrates our handwritten digit recognition/encoding application. 
\footnote{K. Sun, X. Wei, G. Jia, R. Wang, and R. Li are with the team of NerveCloud in the course of EEL-6935: Cloud Computing and Storage, Department of Electrical and Computer Engineering, University of Florida, Gainesville, FL, 32611 USA (e-mail: ksun@ufl.edu). }
\\
\end{abstract}



%
\IEEEpeerreviewmaketitle

\section{Introduction}
\IEEEPARstart{C}{lassification} of data patterns generated by handwriting characters, photograph pixels, audio signals has long been a popular topic in machine learning field. We used to take advantages of the non-volatile feature of computer memory so as to create huge database to afford data searching and pattern matching mechanism. However, we are expecting machines behave more closely to human beings and instead of merely remembering, they could be capable of observing, learning, analysing and recognizing just as human beings' behavior when they come with an unfamiliar object.

Inspired by animal central nervous system, artificial neural network (ANN) comes into the domain of machine learning and pattern recognition \cite{ref_1}. An artificial neural network is made of artificial nodes called ``neurons'', connected together, in order to mimic a biological neural network. One could compare it with human being's brain or neural system. The difference is that each neural node in an artificial neural network not only plays a role of signal carrier, but also contributes to the process of pattern recognition. In other words, the artificial neural network as a whole acts as a human brain.

Inside an artificial neural network, complex global behavior is exhibited by connection between simple processing elements. Neural statistical models consist of sets of weighting factors, which make up two types of connections, a positive weighting factor that has the tendency to increase the activation level of neuron, and a negative one that tends to reduce the output signal of neuron.

\begin{figure}[t]
  
  \centering
    \includegraphics[width=0.4\textwidth]{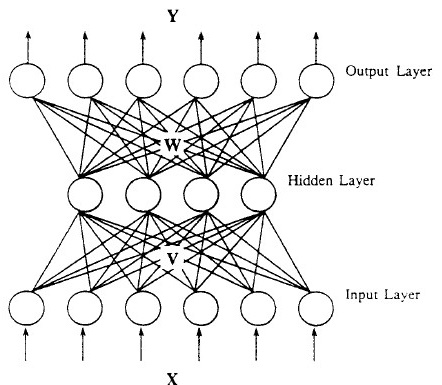}
    \caption{Architecture of ANN \cite{ref_2}.}
  \label{fig:ANN}
\end{figure}

Figure \ref{fig:ANN} shows the basic structure of artificial neural network model. Neurons are located in separated layers, which may contain different numbers of processing elements. Input signals first come to first layer, as known as the input layer, and directly transfer to the middle/hidden layer through weighted connections. The incoming signals are operated by each specific neuron in the hidden layers; in such way output values transfer to each neuron in the output layer through a second layer of weights. At last, via the calculation and operation of the output layer, the output signal is produced. Apparently, the hidden layer may include several layers.

The learning procedures aim at adjusting the weights in the artificial neural network model, so that the performance of models is advanced over time. The learning procedures can roughly be categorized to two types, namely the unsupervised learning and the supervise learning. For unsupervised learning, an input vector from the set of possible network inputs is presented to the network model, then the latter adjusts the weights in order to group the input samples into classes based on their statistical properties. For supervised learning, a set of training samples is presented to the network in sequence. According to the inputs, the network calculates outputs. Via compare between the resulting outputs with an expected output for the particular input sample, some error, which can be used to modify weights, will be found. So the distinguish between unsupervised and supervise learning is that there is no error or feedback to evaluate a potential solution, because the learner is provided with unlabeled samples.

However, the process described above cannot be implemented on any individual computation node because with tremendously increasing number of data on Internet, the incredible complexity of computation would lead to unbounded execution time, thus in turn making the whole mechanism infeasible. Despite the performance, few disks can afford that enormous amount of data, which is usually hundreds of gigabytes or even terabytes. On the other hand, conventional data mining algorithms for classification are not suitable for cloud computing platform either since the weight processing of each layer in back-propagation neural network are dependent on other layers, and in this way, MapReduce is helpless when dividing the entire back-propagation algorithm into mappers and reducers. This problem is carefully reviewed in \cite{ref_old13}. Even though supplied with cluster computing resources, the iteration cycle keeps the same and it can hardly improve the performance of ANN.

In our project, we implement the latest achievement of neural network algorithm, deep learning \cite{deeplearning}, to accelerate the performance of back-propagation neural network. Inspiration comes from the progress we learn to recognize objects. Before studying an unfamiliar object, we are already informed its shape: whether it is a line, a circle, a triangle, or a rectangle and so on. Instead of learning a totally strange object, we learn the representation of the combination of several recognized objects. The fundamental concept of deep learning is that an observation can be represented in multiple ways, but certain representations make it easier to learn tasks of interest from examples. A many-layered neural network could be effectively pre-trained one layer at a time, treating each layer in turn as un unsupervised Restricted Boltzmann Machine, then followed by supervised back-propagation fine-tuning. 

Integrating back-propagation in non-linear deep learning network, our cluster-computing framework is built on top of Amazon Web Service. The aim is quite simple: distribute the complex, large-scale computation to clusters of computers, enforcing parallelized cluster computing.

\section{Related Work}

Of course, we are not the first ones who tried to combine parallel methods of neural network with the cluster computing. Most of existing work achieves parallelization by mapping the nodes in network to the nodes in computing cluster, like \cite{ ref_old9, ref_old10, ref_old11}. However, these algorithms are not suitable for MapReduce. In MapReduce structure, users cannot specifically control a certain node in a cluster. Instead, MapReduce can only assign the mapper's and reducers' jobs as a whole, so the algorithms above are not implementable in this cloud computing environment. Moreover, the scale of these work is relatively small, and lack of elasticity. Mapping the network nodes to different computing nodes will inevitably increase the I/O cost. In most cases, neural network problem are dealing with big data, which requires large amount of I/O operations. As the result, I/O cost is the major cost in the distributed computing environment of existing methods.

Chu et al. \cite{ref_old12} and Liu et al. \cite{ref_3} both proposed their own algorithm that used back-propagation algorithm to train a three-layer neural network based on MapReduce. They did the supervised learning (which is the only kind of learning back-propagation can do) to classify the input data into two categories and did some experiments on multi-core environment. However, They did not adopt the latest neural network achievement, deep learning, and therefore still suffered from the problem of low learning efficiency.

\section{System Architecture}

\subsection{Main Strategy}

In detail, we implement deep learning algorithm \cite{deeplearning} to train the input data, where a MapReduce programming model is made use of to parallelize the computation. The MapReduce job consists of the mapper and reducer functions, of which the mapper function will extract key/value pairs from input data and transfer them into a list of intermediate key/value pairs, and reducer function merges these intermediate values corresponding to the same key generated from mapper function to produce output values. In the case of Machine Learning, the input value will be the data from a certain object that a machine is going to ``learn'', and in our project, the objects are data sets extracted from handwriting characters. The intermediate key/value pair will be the weights, in order for a machine to determine whether it has acknowledged the object correctly. Reducer function uses these weights to compute the so-called ``acknowledgment'' of the machine of an intended object \cite{ref_4}. If there exits an intolerable difference between the precision of training set and expected precision, the MapReduce job will loop until an acceptable result has worked out.

\subsubsection{Difficulties}

As mentioned before, the main problem is the huge amount of data waiting to be trained. Although deep learning learns from representation, it is not capable of eliminating similarity and noisiness contained in data sets, and this will terribly affect machine learning \cite{ref_5}. Thus the most straightforward consequence is that the MapReduce job would loop many times so that it is hard to provide a satisfying precision, or even it cannot produce an output. It is necessary, however, to implement a mechanism to somehow improve the efficiency of the procedure of neural network. One idea is to remove the similar items by using diversity-based data sampling method. It can be applied to MapReduce programming model, too, where the frequency of input data will be counted, and those duplicated data is eliminated. Another way is to somehow train the data so that the machine would find a certain pattern corresponded with the data set. Both are applicable to develop the efficiency of back-propagation.

\subsection{Modules and Subsystems}

\begin{figure}[t]
  \centering
    \includegraphics[width=0.5\textwidth]{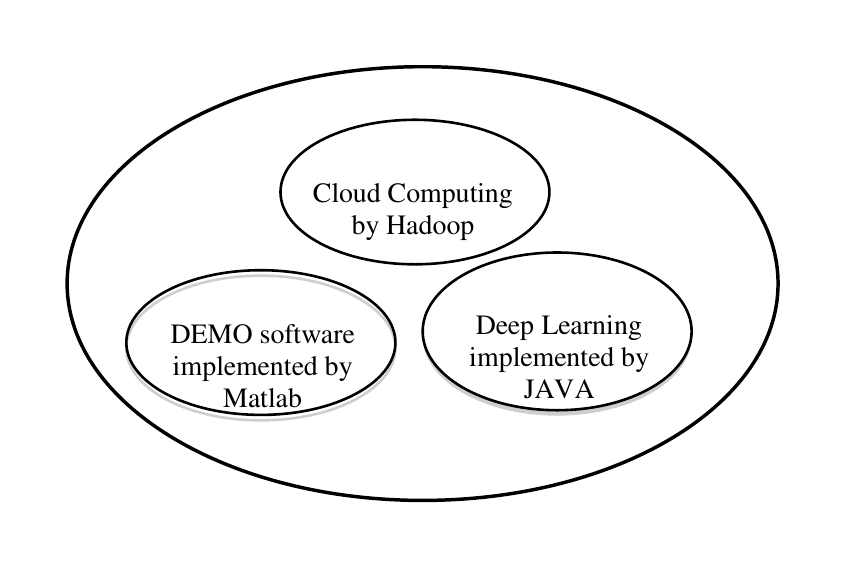}
    \caption{Modules and Subsystems of the Design.}
  \label{fig:subsystems}
\end{figure}

Our system is made up of three different parts, the Deep learning implemented by Java, the Cloud Computing by Hadoop, and Demonstration software implanted by Matlab. Such design can be well illustrated in figure \ref{fig:subsystems}.

Briefly, we divide our progress of machine learning into three steps: the step of pre-training which makes use of deep learning technology to initialize weights, the step to fine-tune the weights, and the last step aiming at improving the precision.

\section{Deep Learning}

\subsection{Pre-training}

The motivation to insert pre-training step before fine-tuning is the unacceptable inefficiency of back propagation when converting a high-dimensional data into low-dimensional codes. When we have multiple layers, large initial weights always witnesses poor local minima, while small initial weights will lead to tiny gradients in early layers, making it infeasible to train the whole back propagation system with many hidden layers \cite{ref_6}.

Therefore, we have to make the initial weights close to the solution. The idea is simple. Think about two pictures of number zero and number one. Instead of jumping to the final step of recognizing the number, we let machine accomplish the task step by step, where it first looks for the internal pattern of two pictures, namely, recognizing a circle and a stick, then encodes the circle and the stick to zero and one respectively. Practically, however, it is not easy because this process requires a very different type of algorithm that learns one layer of features at a time. 

\begin{figure}[]
  \centering
    \includegraphics[width=0.5\textwidth]{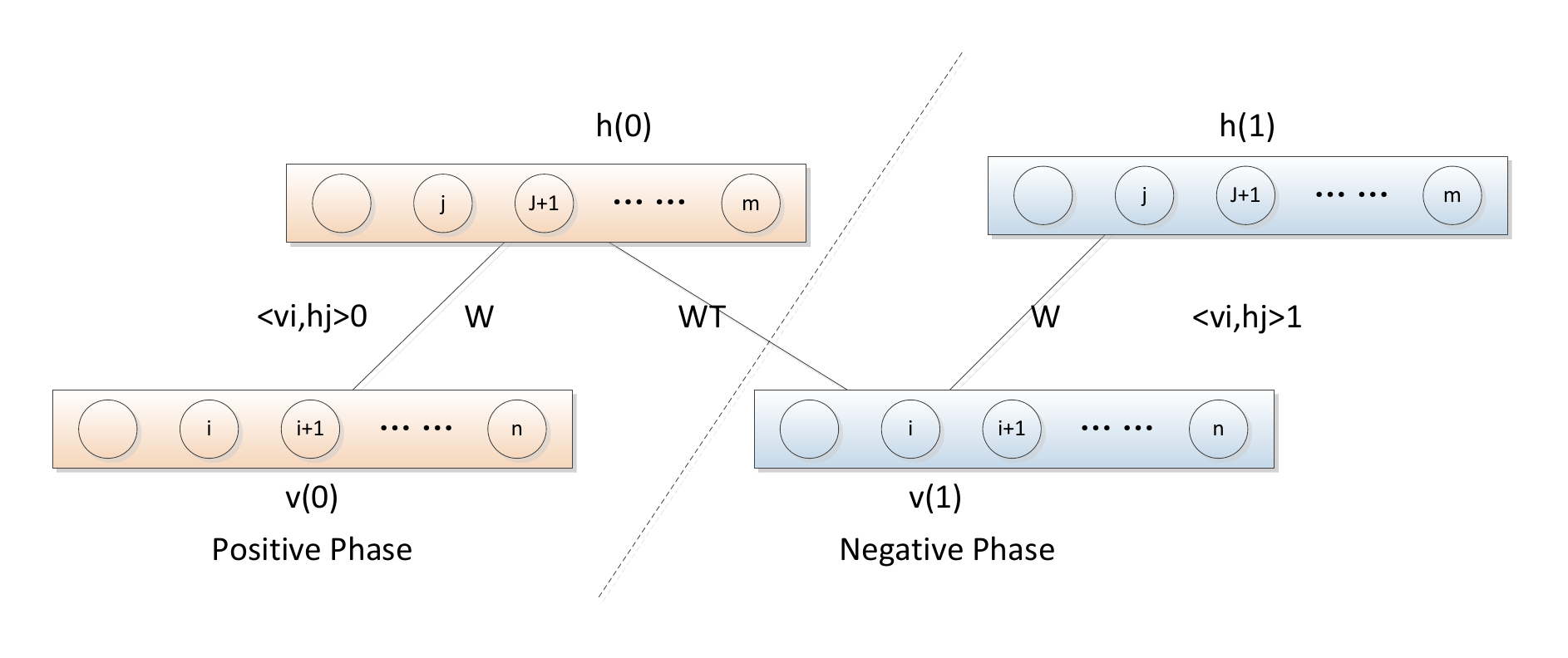}
    \caption{Algorithm Flow Char for Restricted Boltzmann Machine (RBM).}
  \label{fig:RBM}
\end{figure}

We make advantage of restricted Boltzmann machine (RBM) to realize pre-training, where an ensemble of binary vectors can be modelled using a two-layer network, in which stochastic, binary pixels are connected to stochastic, binary feature detectors using symmetrically weighted connections. Pre-training involves learning a stack of RBMs, each of which has one layer of feature detectors, and it is sort of recursive process where the learned feature activations of one RBM are used as ``data'' for training the next RBM in the stack. For each RBM, we are not going to evaluate the weights between two layers for infinite times in order for the efficiency. Instead we get the modified weights result from two rounds of RBM calculation, and the result of this recursive execution is unrolled to create a deep auto encoder. Note that this does not hurt the precision quite much because of the mathematics proof by Geoffrey E. Hinton. Thanks to the largely increased number of machines involved in, the time cost for training will decrease in inverse proportion, too. Figure \ref{fig:RBM} give us a directly illustration of what RBM is.

One more point to be noticed is that we are adjusting the weights of paths based on the average variations of a single weight resulting from the modification made by a batch of training items. The reason is updating the value of weights once a single training item is extremely slow and inapplicable for MapReduce method.

\subsection{Fine-tuning}

This is the step where we get zero and one from a circle and a stick. Pre-training initializes the weights so that they are close to the solution we want, but they are not the answer.
We train the weights using back propagation \cite{ref_7} algorithm based on MapReduce method in multiple layer neural network \cite{ref_8}. Back propagation and its improved methods are applied in
mobile data processing. Such applications can be found in
\cite{ref_9, ref_10, ref_11}. Note that here the trained data is not randomized weights but well-initialized weights, making it easier to get the solution. Every mapper receives one training item and then computes all update value of the weights. Then each reducer gathers update-values for one weight and calculating the average. The input value of mapper is the input item while the input key is empty. On the other hand, the value of the reducer is the difference between original weight and updated weight, and the corresponding key is the weight. Similar to pre-training, variations of each weight should be updated batch by batch for the sake of efficiency.

\subsection{Precision Refinement}
This step is still in discussion. We are going to use Adaboosting \cite{ref_12} method to refine the result of data training. Basically, a neural network model after training is regarded as a classifier. Unfortunately, the precision of a classifier is vulnerable to the noisy and the large-scale of the mobile data. Such learning is called weak learning, and we need to improve the performance of the classifier.

The general concepts of the Adaboosting method can be summarized as: get one weak classifier from part of the training set; get more using different parts of the training set sampled out by the features of the former one; assemble them. This is just the basis and we are not going to talk much about the Adaboosting method here because we do not have a good idea about it now. Our main focus at present is implementing the pre-training step, using deep learning, and realizing the MapReduce based back-propagation algorithm.

\section{Cloud Computing based on MapReduce}

\subsection{Choice of Cloud Computing Platform}

One of the most severe disadvantages of deep learning is the long training time. Such time-consuming problem prohibit trained machine from quickly equipped with accurate nueral network data and accomplish required task. To solve such problem, applying cloud computing algorithm to machine learning is a good idea.  

Among all the MapReduce platform, Hadoop \cite{Hadoop} is a good choice for our project. It is an open source software for data storage as well as large scale processing of data-sets on clusters of commodity hardware \cite{ref_13}. The main modules that compose Hadoop framework are Hadoop Common, Hadoop Distributed File System (HDFS), Hadoop Yarn,and Hadoop MapReduce \cite{ref_14}. Hadoop Common contains all the utilities and libraries that required by all the other Hadoop modules. HDFS is a distributed file-system that containing data on the clustering machines. It can provide highly aggregated bandwidth for both masters and slaves across the commodity. Hadoop Yarn is a resource-management platform. It is responsible for arranging computing resources in commodity through which the users' applications are scheduled. Hadoop MapReduce is the most important part here. It provides a programming model for large scale data. 

In this project, we use Amazon Web Service EC2 platform to achieve our MapReduce algorithm. It is a collection of remote computing services that provides a cloud computing platform. Based on a physical srver farm, it can provides customer faster and cheaper large computing capacity.

\subsection{MapRedeuce Structure Design}

\begin{figure}[]
  \centering
    \includegraphics[width=0.5\textwidth]{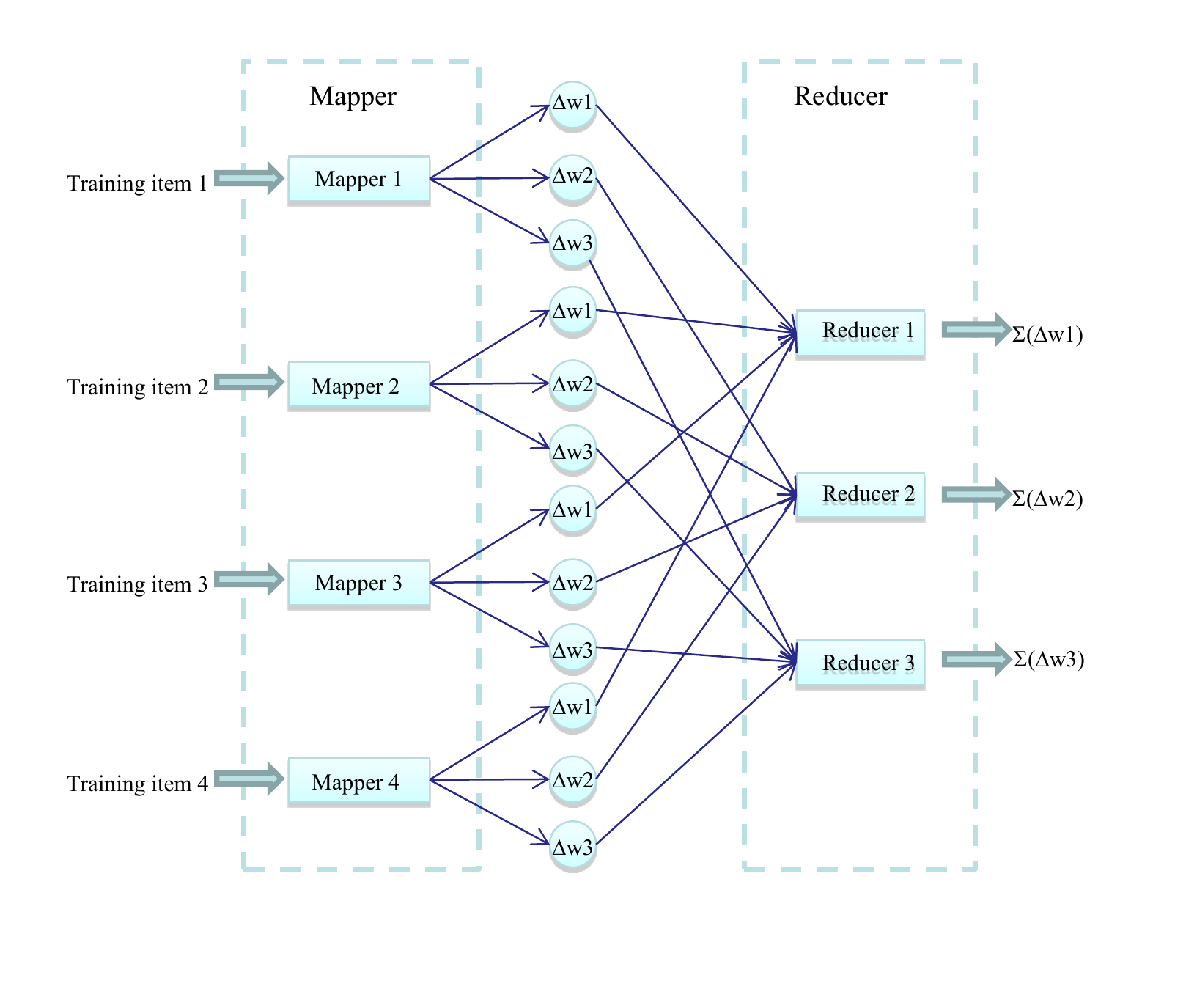}
    \caption{Flow chart of applying MapReduce on deep learning.}
  \label{fig:MapReduce}
\end{figure}

In order to implement MapReduce to RBM, we apply an algorithm that can be well illustrated in figure \ref{fig:MapReduce}. For the sake of easy calculation, all the weights coming from paths between nodes are allocated in a matrix. For each mapper task, one training item is sent to a mapper. Their outputs are matrices of variables of weight resulted through the RBM algorithm. A unique ID can identify every element in the output matrix of mapping tasks, which is considered as the key of the reducing task. For each reducer task, a reducer accepts the ID and the value of a certain element of the matrix as key value pair. Since each reducer will only receive variable value of one particular ID, it can sum up all the results provided by each training item and find the final update of that element.

\subsection{Pseudo-code for Algorithm}

Our MapReduce-based deep learning algorithm contains 6 Java classes. A MapReduce driver class ($DeepLearningDriver.java$), two mappers and two reducers for RBM training and forward propagation tasks ($RBMMapper.java$, $RBMReducer.java$, $PropMapper.java$, and $PropReducer.java$), and a class that implements all the matrix operations ($Matrix.java$).

We present our algorithms using Pseudo-codes listed below. First, Algorithm \ref{alg:Driver} describes the driver of deep learning MapReduce program. It contains two MapReduce structures: RBM and forward propagation. It is the scheduler for the MapReduce tasks.

\begin{algorithm}[]     
\caption{ The Driver for MapReduce-based Deep Learning algorithm }
\label{alg:Driver}
\begin{algorithmic}[1] 
\REQUIRE ~~\\
  The user provides the input file location, output file location, max iteration number ($maxEpoch$), number of layers ($numLayer$), number of nodes in each layer ($numNodes(layer)$) via input arguments.
\ENSURE ~~\\
\FOR    {$layer \leftarrow 1$ to $numLayers - 1$}
	\STATE $numVis \leftarrow numNodes (layer - 1)$;
	\STATE $numHid \leftarrow numNodes (layer)$;
	\STATE $Weights(layer) \leftarrow RandomizeWeights()$;
	\STATE $iter \leftarrow 1$;
	\FOR {$iter \leq$ Max Iteration Times}
		\STATE $Job_{RBM} \leftarrow$ Initialized MapReduce Job for RBM;
		\STATE Store $Weights(layer)$ into  $\rightarrow$ file system: $FS$;
		\STATE Give config data: $numVis \rightarrow Job_{RBM}$;
		\STATE Give config data: $numHid \rightarrow Job_{RBM}$;
		\STATE Assign Distributed Cache: $FS \rightarrow Job_{RBM}$; \\
		// Each  Mapper has a full copy of weights
		\STATE Start $Job_{RBM}$ :
		\STATE input: the network input for current layer; \\ 
		// (\# of Mappers: num of cases)
		\STATE output: $Weights_{update} \leftarrow$ update for every weight; \\
		// (\# of Reducers: $numVis \times numHid$)
		\STATE  Update $Weights(layer)$:
		\FORALL {$weight \in Weights(layer)$ }
			\STATE $weight \leftarrow weight + weight_{update}$;
		\ENDFOR
		\STATE $iter$++;
	\ENDFOR
	\STATE $Job_{prop} \leftarrow$ Initialized MapReduce Job for propagate;
	\STATE Store $Weights(current)$ into  $\rightarrow$ file system: $FS$;
	\STATE Give config data: $numVis \rightarrow Job_{prop}$;
	\STATE Give config data: $numHid \rightarrow Job_{prop}$;
	\STATE Assign Distributed Cache: $FS \rightarrow Job_{prop}$; 
	\STATE Start $Job_{prop}$ :
	\STATE input: the network input for current layer; 
	\STATE output: the network output for current layer;
\ENDFOR 
\RETURN ~~\\
  The final $Weights(last\_layer)$ is the trained result we want.
\end{algorithmic}
\end{algorithm}

Algorithm \ref{alg:RBMMapper} describes the mapper of restricted Boltzmann machine (RBM) training part in deep learning MapReduce program. Each mapper only trains the weights for one iteration using one test case. So, in order to train the weights of the whole neural network, the MapReduce program needs to execute $(maxEpoch \times numLayers)$ times. It contains six parts: $configure()$ reads all the configurations and distributed cache from outside; $initialize()$ parse the input strings into parameters, and initialize parameters for algorithm; $getposphase()$ does the positive phase of RBM training; $getnegphase()$ does the negative phase of RBM training; $update()$ computes the update of weights using previous results and parameters; $map()$ implements the mapper. It outputs the original key and updated value pair as the intermediate data. In the pseudo-code, we skip the details of RBM training algorithm, because that is not the focus of this paper. If the reader is interested in it, please refer to Dr. Hinton's paper in \cite{deeplearning}.

\begin{algorithm}[]     
\caption{ The mapper of restricted Boltzmann machine (RBM) training part in deep learning MapReduce program. }
\label{alg:RBMMapper}
\begin{algorithmic}[1] 
\REQUIRE ~~\\
  Input of the mapper is one training case from network input. Also, there are arguments like $numVis$, $numHid$ and $Weight(current)$ past in via configurations or distributed cache.
\ENSURE ~~\\
\STATE $initialize()$;
\STATE $getposphase()$;
\STATE $getnegphase()$;
\STATE $update()$;
\FOR {$i \leftarrow 0$ to $numVis - 1$}
    \FOR {$j \leftarrow 0$ to $numHid - 1$}
        \STATE output $\langle key, value\rangle$ pair:  \\ $\langle Weight_{ID}, Weight_{Update}\rangle$
    \ENDFOR
\ENDFOR
\RETURN ~~\\
  Each mapper output its update of $Weights()$ according to its train case.
\end{algorithmic}
\end{algorithm}

Algorithm \ref{alg:RBMReducer} describes the reducer of restricted Boltzmann machine (RBM) training part in deep learning MapReduce program. Each reducer collects all the weight updates from one single weight ID. It adds up the updates from the same weight and write to the final output.

\begin{algorithm}[]     
\caption{ The reducer of restricted Boltzmann machine (RBM) training part in deep learning MapReduce program. }
\label{alg:RBMReducer}
\begin{algorithmic}[1] 
\REQUIRE ~~\\
  Input of the reducer is the intermediate data output by the mappers from the same weight ID.
\ENSURE ~~\\
\STATE $sum \leftarrow 0$;
\FORALL {$Weight_{update} \in$ same $Weight_{ID}$}
    \STATE $sum \leftarrow sum +Weight_{update}  $;
\ENDFOR
\STATE output $\langle key, value\rangle$ pair:  \\ $\langle Weight_{ID}, sum\rangle$
\RETURN ~~\\
  Each reducer output the overall update of $Weights()$.
\end{algorithmic}
\end{algorithm}

Algorithm \ref{alg:PropMapper} describes the mapper of forward propagation part in deep learning MapReduce program. It is executed between every two layers. So the total execution time for this MapReduce program is $numLayer - 1$. It contains four parts: $configure()$ reads all the configurations and distributed cache from outside; $initialize()$ parses the input strings into parameters, and initializes parameters for algorithm; $prop2nextLayer()$ computes the forward propagation algorithm; $map()$ implements the mapper. It outputs the original key and updated value pair. We also skip the algorithm details here, and just focus on the structure.

\begin{algorithm}[]     
\caption{ The mapper of forward propagation part in deep learning MapReduce program. }
\label{alg:PropMapper}
\begin{algorithmic}[1] 
\REQUIRE ~~\\
  Input of the mapper is one training case from network input. Also, there are arguments like $numVis$, $numHid$ and $Weight(current)$ past in via configurations or distributed cache.
\ENSURE ~~\\
\STATE $initialize()$;
\STATE $prop2nextLayer()$;
\STATE Initialize string $update \leftarrow $ ``''
\FOR {$i \leftarrow 0$ to $numHid - 1$}
    \STATE $update \leftarrow update + Case_{update} $ + `` '';
\ENDFOR
\STATE output $\langle key, value\rangle$ pair:  \\ $\langle Case_{ID}, update\rangle$
\RETURN ~~\\
  Each mapper output the forward-propagated training case $Case_{update}$.
\end{algorithmic}
\end{algorithm}

The reducer of forward propagation part is just output the intermediate data as final output, so it is omitted.

\section{Performance and Results}

In the performance evaluation section, extensive experiments have been conducted. The experimental results prove the efficiency, and scalability of our proposed MapReduce on neural network method. This method can be applied over large-scale realistic data on the cloud-computing platform of AWS.

Because the goals of our project consist of two parts: deep learning on neural network and MapReduce-based parallelized computing, three sets of experiments have been conducted: objective performance evaluation of deep learning algorithm, experiments on speed-up gained from MapReduce, and subjective performance evaluation of demo applications. 

\subsection{Performance Experiment Set-up}

We use the hand-written digit training and testing cases from the on-line database available in \cite{digit}. It has a training set of 60,000 examples, and a test set of 10,000 examples. And all of them are labeled.

The AWS cloud-computing platform is built up by multiple EC2 instances. Up to 32 EC2 nodes have been used in these experiments for comparing the performance of experiments in different running instances. Massive input data and intermediate results are stored in the distributed cache offered by the platform. The raw input data is in the size of 300 megabytes and stored distributed across the platform. Each EC2 instances has Intel 64 bit CPU, 16 GB memories and high network performance. Moreover each node can be boosted up to 4 virtual CPU with the performance triple increased. Open source framework Hadoop is adopted by AWS for the distributed architecture of those nodes.

\begin{figure}[]
  \centering
    \includegraphics[width=0.4\textwidth]{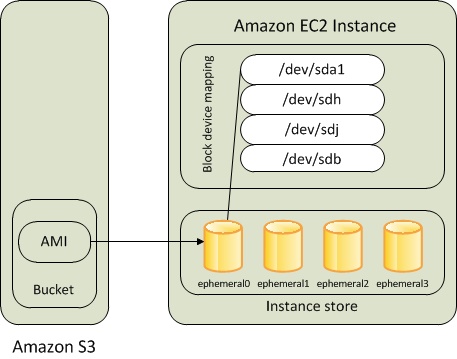}
    \caption{Amazon EC2 instances architecture.}
  \label{fig:AWS}
\end{figure}

Amazon EC2 instances provide a number of additional features to deploy, manage, and scale our applications. Multiple storage options based on our requirements can be choose. Details about how EC2 instances work can be found in figure \ref{fig:AWS}.

\subsection{Result Evaluations}

\begin{figure}[h]
\centering
    \subfigure[Training Error v.s. iteration time curve]
    {
        \includegraphics[width= 0.225\textwidth]{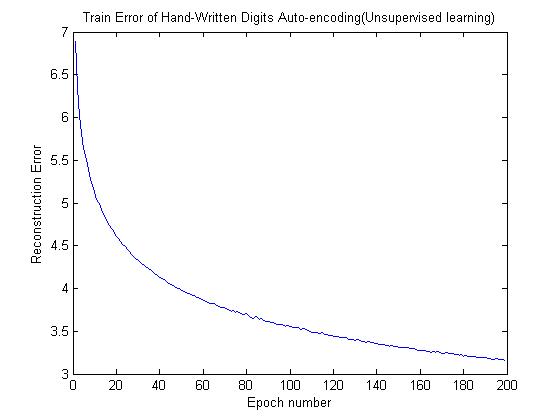}
    }
    \subfigure[Testing Error v.s. iteration time curve]
    {
        \includegraphics[width=0.225\textwidth]{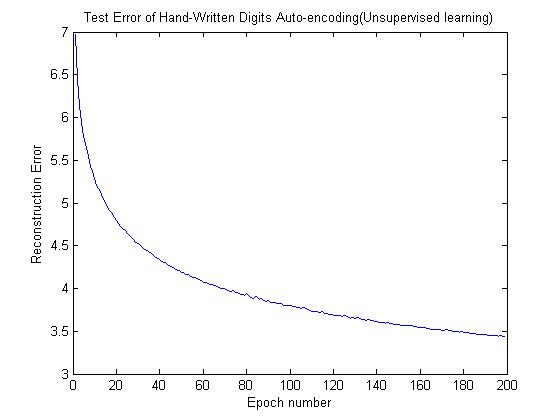}
    }
    \caption{Train and test error v.s. iteration time curve for unsupervised learning.} 
    \label{fig:auto-result}
\end{figure}

In the first part, we conduct the objective performance evaluation of deep learning algorithm. Error rate of our system is the primary concern, therefore we conducting several experiments to measure the error rate after several iterations. Figure \ref{fig:auto-result} shows the number of training error and testing error as the iteration number increases, for unsupervised learning of hand-written digits. As we can see from figure \ref{fig:auto-result}, the error rate for reconstruction of hand-written digits decreases significantly after several iterations, for both training cases and testing cases. This is achieved by the application of RBM algorithm and back-propagation algorithm. 

\begin{figure}[h]
\centering
    \subfigure[Training Error v.s. iteration time curve]
    {
        \includegraphics[width= 0.225\textwidth]{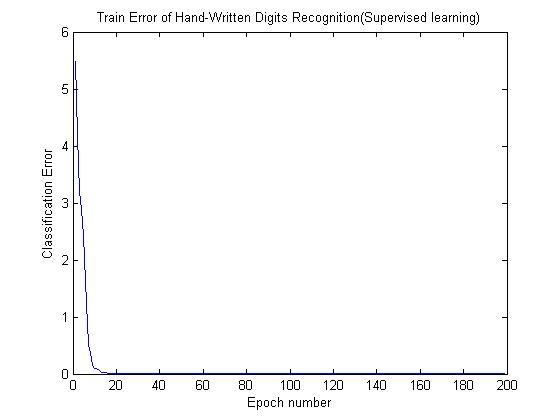}
    }
    \subfigure[Testing Error v.s. iteration time curve]
    {
        \includegraphics[width=0.225\textwidth]{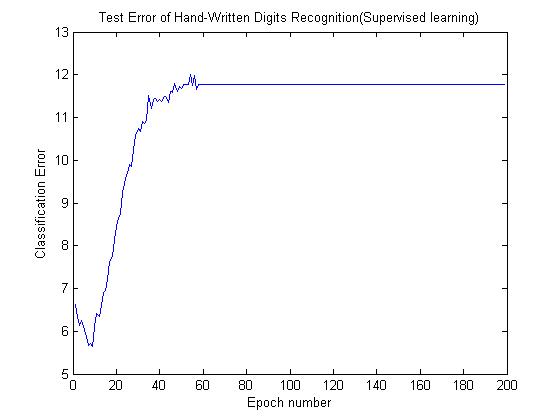}
    }
    \caption{Train and test error v.s. iteration time for supervised learning. Note that the misclassification decreases in training (a), but increases in testing (b), which is a sign of over-fitting problem. } 
    \label{fig:class-result}
\end{figure}

Figure \ref{fig:class-result} shows the number of misclassification in training and testing as the number of iterations increases, for supervised learning of hand-written digits recognition. As we can see in figure \ref{fig:class-result} (a), the training error soon reaches 0 after several iterations. However, in figure \ref{fig:class-result} (b), it shows that the misclassification rate of testing cases increases as number of iterations increases. This is a sign of over-fitting problem, which is clearly discussed in \cite{overfit}. Over-fitting is inevitable in supervised learning and classification problems.

\begin{figure}[]
  \centering
    \includegraphics[width=0.4\textwidth]{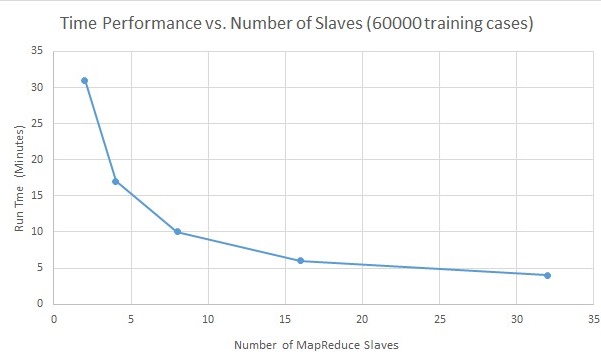}
    \caption{Performance result -- time performance v.s. number of slaves curve for unsupervised learning.}
  \label{fig:performance}
\end{figure}

In the second part, we conduct the experiments on speed-up gained from MapReduce. The efficiency and scalability of our MapReduce method is tested on cloud clusters with 2,4,8,16,and 32 nodes. The test input size is 300 MB. From our result, we can clearly conclude that the running time of tasks is linearly dependent on the aspects of nodes numbers in the cluster. The slightly mismatch between result and our anticipation is because of the overhead of system architecture. Our result proves MapReduce have an excellent speed-up performance. Figure \ref{fig:performance} shows the running time has an inverse ratio relationship with the nodes number approximately. Our result proves our method has an excellent scalability.

\subsection{GUI Demo by MATLAB}

To further refine our project, we implement two GUI demo software inherited from MATLAB GUI package. GUIs also known as graphical user interfaces will provide a type of point-and-click control of the software application we designed. By implementing a GUI interface, users are easier to understand the performance of our system. The reason why we choose to use MATLAB GUIs is MATLAB apps are self-contained programs with GUI front ends that automate a test. MATLAB GUIs usually includes controls such as menus, toolbars, buttons, and sliders.

\begin{figure*}[t]
\centering
    \subfigure[import the weight file for classification]
    {
        \includegraphics[width= 0.3\textwidth]{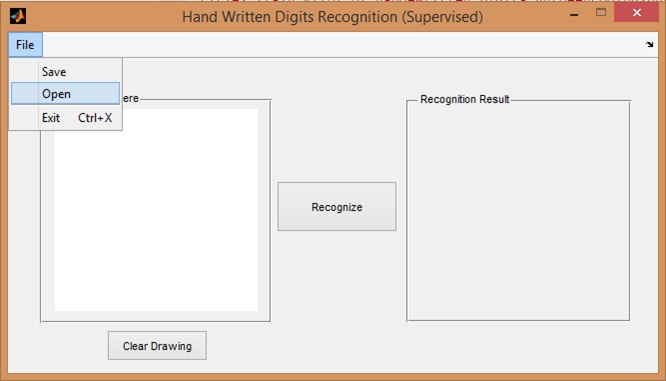}
    }
    \subfigure[draw a digit with mouse]
    {
        \includegraphics[width=0.3\textwidth]{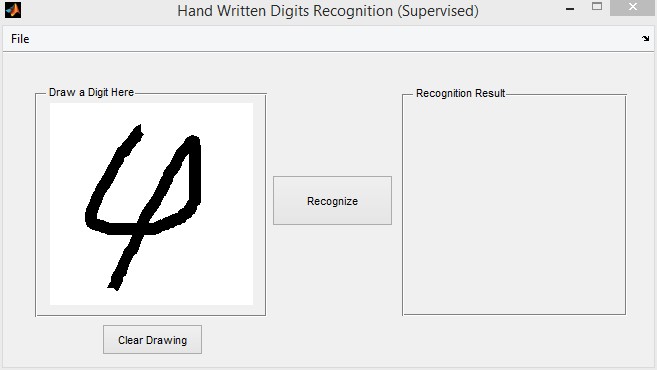}
    }
    \subfigure[click recognize button and the result shows]
    {
        \includegraphics[width=0.3\textwidth]{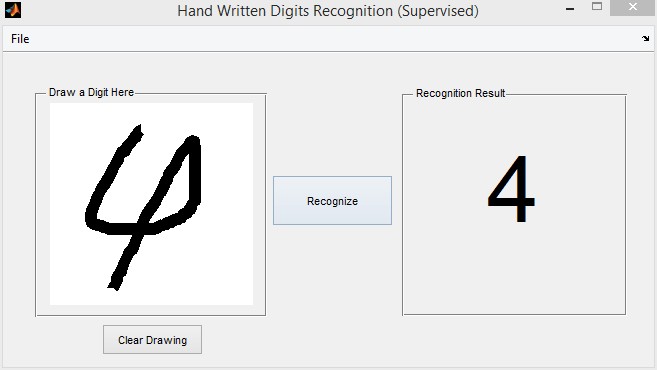}
    }
    \caption{Screen shots of demo program for supervised learning of hand-written digit recognition. (a) step is to import the weight file; (b) step is to draw a digit with mouse on left box; (c) step is to click the recognize button and the result shows on the right.} 
    \label{fig:supervised-demo}
\end{figure*}

\begin{figure*}[t]
\centering
    \subfigure[import the weight file for auto-encoding]
    {
        \includegraphics[width= 0.225\textwidth]{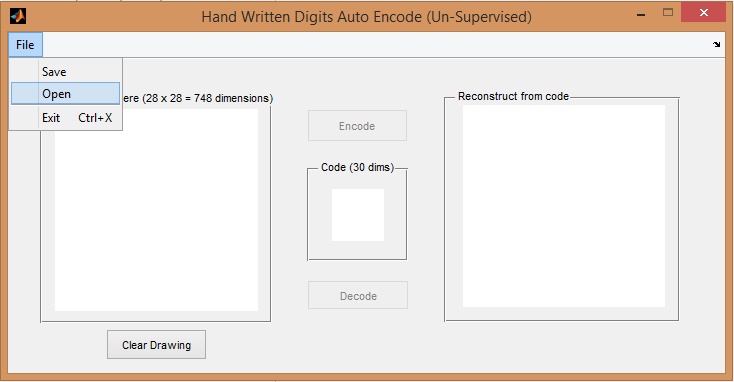}
    }
    \subfigure[draw a digit with mouse]
    {
        \includegraphics[width=0.225\textwidth]{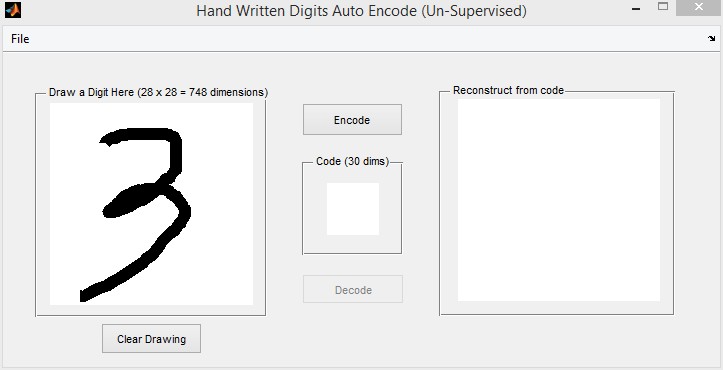}
    }
    \subfigure[click encode button and the code shows in the middle]
    {
        \includegraphics[width=0.225\textwidth]{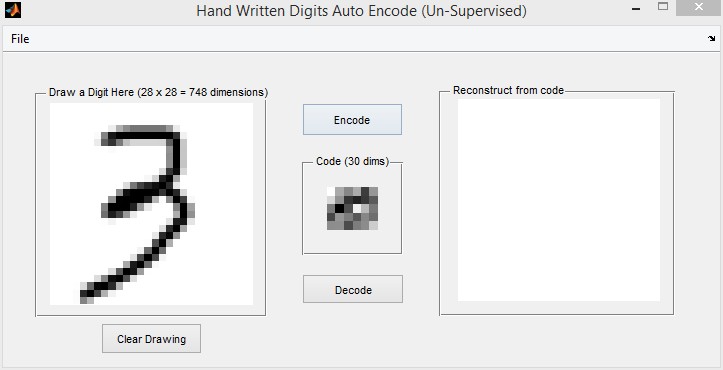}
    }
    \subfigure[click decode button and the reconstruction shows on the right]
    {
        \includegraphics[width=0.225\textwidth]{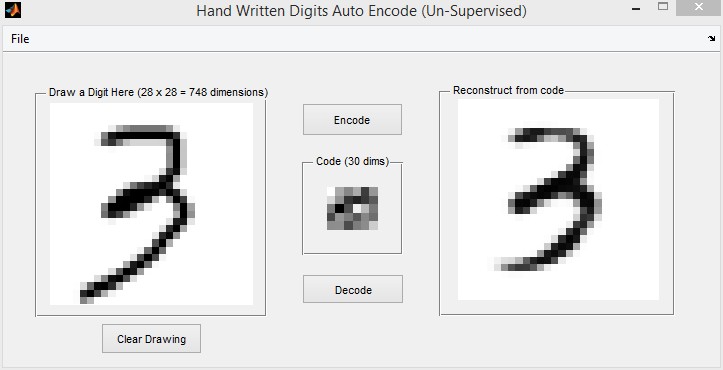}
    }
    \caption{Screen shots of demo program for unsupervised learning of hand-written digit auto-encoding. (a) step is to import the weight file; (b) step is to draw a digit with mouse on left box; (c) step is to click encode button and the code shows in the middle; (d) step is to click decode button and the reconstruction shows on the right.} 
    \label{fig:unsupervised-demo}
\end{figure*}

There are two demos: figure \ref{fig:supervised-demo} shows the screen shots of demo software for supervised learning (hand-written digit recognition) results, and figure \ref{fig:unsupervised-demo} shows the screen shots of the demo software for unsupervised learning (hand-written digit auto-encoding and auto-decoding) results. 

As figure \ref{fig:supervised-demo} shown, to use the software, there are three steps:(a) step is to import the corresponding supervised learning weights file by selecting ``File $\rightarrow$ Open''; (b) step is to draw a digit with mouse on left box; (c) step is to click the ``recognize'' button and the recognition result shows on the right.

As figure \ref{fig:unsupervised-demo} shown, to use the software, there are four steps:(a) step is to import the corresponding unsupervised learning weights file by selecting ``File $\rightarrow$ Open''; (b) step is to draw a digit with mouse on left box; (c) step is to click encode button and the code shows in the middle; (d) step is to click decode button and the reconstruction shows on the right. Note that the picture is $28 \times 28$ dimensions, and the code in the middle only has 30 dimensions, so the compress rate is $30 \div 784 = 0.357$. 

In the last part, we present a subjective evaluation of our system using some screen shots of our demo software. 

\begin{figure}[]
\centering
    \subfigure[]
    {
        \includegraphics[width= 0.225\textwidth]{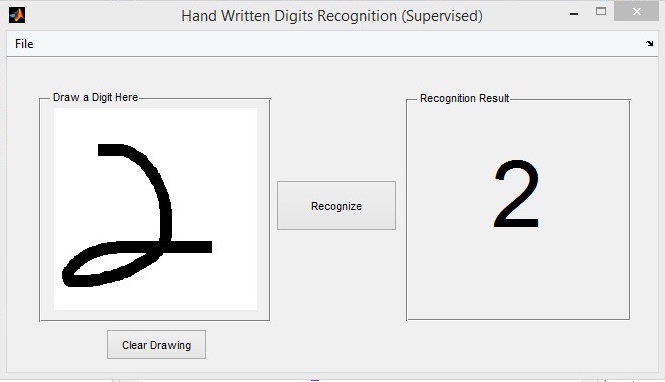}
    }
    \subfigure[]
    {
        \includegraphics[width=0.225\textwidth]{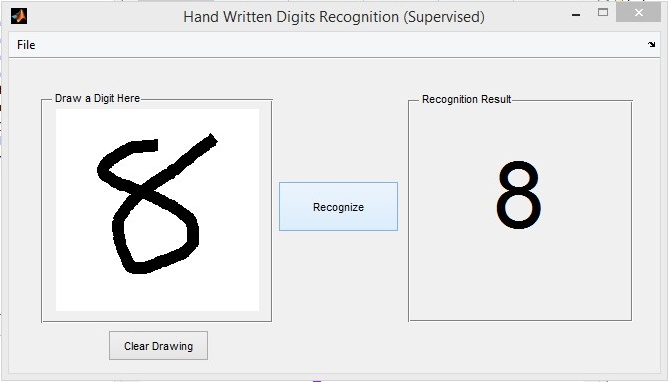}
    }
    \subfigure[]
    {
        \includegraphics[width=0.225\textwidth]{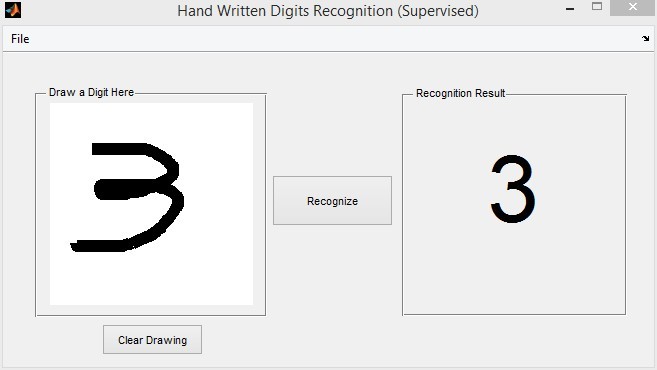}
    }
    \subfigure[]
    {
        \includegraphics[width=0.225\textwidth]{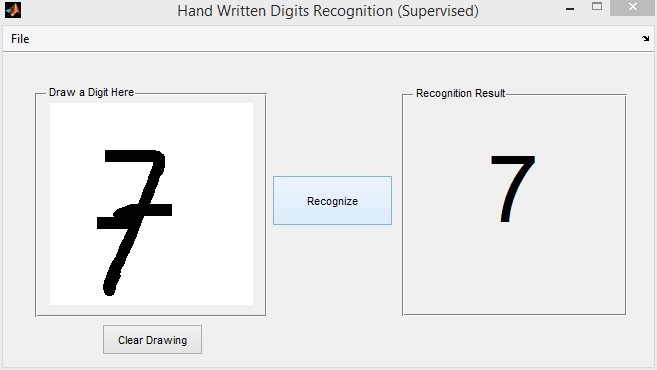}
    }
    \caption{Demo screen shots of supervised learning of hand-written digit recognition. In every figure, the left side contains the hand-written digit written by user using mouse, and the right side is the recognition result.} 
    \label{fig:supervised-results}
\end{figure}

The supervised learning (hand-written digit recognition) results are shown in figure \ref{fig:supervised-results}. As we can see, the recognition results are very accurate, despite of some informal handwriting, like 7 with a bar.

\begin{figure}[]
\centering
    \subfigure[]
    {
        \includegraphics[width= 0.225\textwidth]{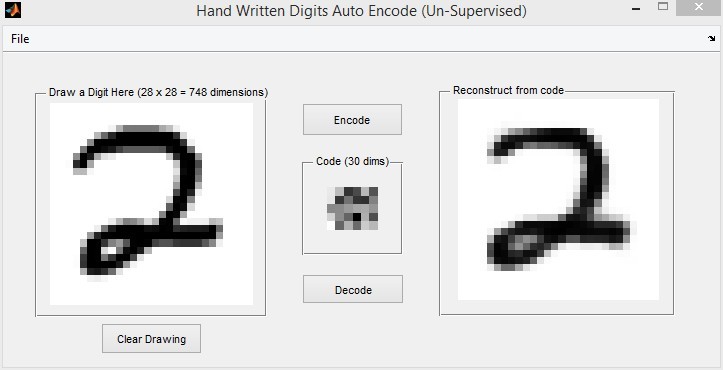}
    }
    \subfigure[]
    {
        \includegraphics[width=0.225\textwidth]{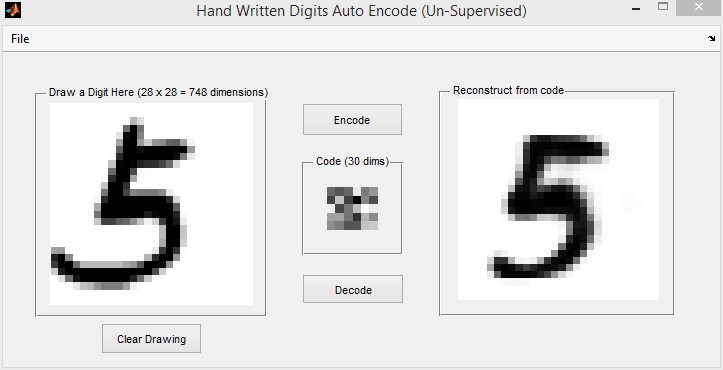}
    }
    \subfigure[]
    {
        \includegraphics[width=0.225\textwidth]{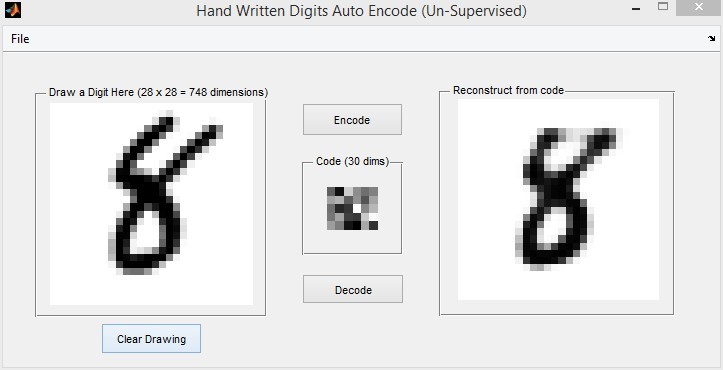}
    }
    \subfigure[]
    {
        \includegraphics[width=0.225\textwidth]{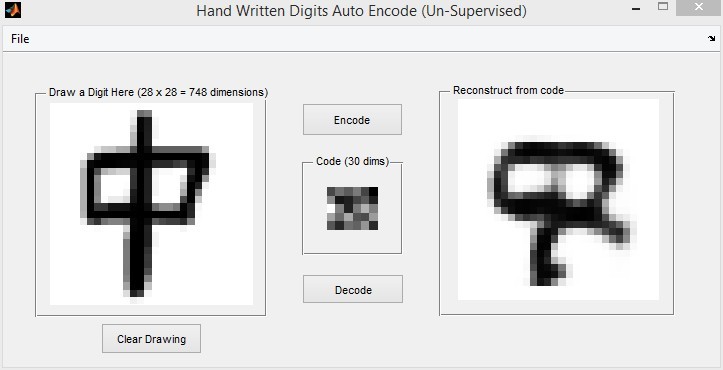}
    }
    \caption{Demo screen shots of unsupervised learning of hand-written digit auto-encoding. In every figure, the left side contains the hand-written digit written by user using mouse (down-sampled as $28 \times 28$ pixels), the middle small box contains the code for left picture (only 30 dimensions), and the right side is the auto-reconstruction result from the code in the middle. Note the in (a) (b) and (c), the reconstruction is very similar to the original picture, but in (d), they are quite different.} 
    \label{fig:unsupervised-results}
\end{figure}

The unsupervised learning (hand-written digit auto-encoding and auto-decoding) results are shown in figure \ref{fig:unsupervised-results}. In figure \ref{fig:unsupervised-results} (a - c), the numbers are nicely reconstructed after encoding and decoding. Bear in mind that the recovery is purely based on the code in the middle, but not the original picture. Although there is very subtle difference between them, you can definitely understand the reconstructed number. 

However, if we write a different category of symbols (like a Chinese character) in the box on the left, instead on digits, the reconstruction ability is very poor. Moreover, we can see from figure \ref{fig:unsupervised-results} (d), the program ``tries'' to recover the symbol into a digit. That is because we trained the neural network with hand-written digits, and the network does not recognize other symbols. 

\section{Conclusions}

In general, we successfully designed and implemented a MapReduce neural network algorithm on large scale of data which running on top of Amazon Web Service platform. Great improvements on efficiency and accuracy have been achieved by our system, due to the facts that our system is running on distributed file platform. We also observe that the running time of processing data is decreasing as the number of nodes increasing, almost in an inverse linear mode. The slightly mismatch between experiments and theory is because of the system overhead when larger number of nodes is introduced into the system. We are convinced that with the application of MapReduce, neural network is capable of successfully recognizing handwriting digits with great efficiency and accuracy. In the future we are planning to apply our algorithm on more complex input model including face recognition, speech recognition and human language processing.

\section*{Appendix}

The code repository can be found on Github at: \url{https://github.com/sunkairan/MapReduce-Based-Deep-Learning}

\section*{Acknowledgement}

The authors thank Dr. Andy Li and his teaching assistances for their suggestions that significantly improved the quality of the paper. The authors also thank Dr. Dapeng Oliver Wu and his team for providing deep discussions about our research details. The computing cluster resources used by this project is supported by the course of EEL-6935: Cloud Computing and Storage.



%


\bibliographystyle{IEEEtran}
\bibliography{IEEEabrv,DLMR}

\end{document}